\documentclass[twocolumn, tighten]{aastex63}

\def\lum{erg~s$^{-1}$}

\newcommand{\chandra}{{\itshape Chandra}}

\newcommand{\nustar}{NuSTAR}

\newcommand{\xmm}{XMM-Newton}

\def\xray{\hbox{X-ray}}

\def\ltsima{$\; \buildrel < \over \sim \;$}
\def\simlt{\lower.5ex\hbox{\ltsima}}
\def\gtsima{$\; \buildrel > \over \sim \;$}
\def\simgt{\lower.5ex\hbox{\gtsima}}
\def\kms{\ifmmode{~{\rm km~s^{-1}}}\else{~km s$^{-1}$}\fi}
\def\lsim{\lower0.3em\hbox{$\,\buildrel <\over\sim\,$}}
\def\gsim{\lower0.3em\hbox{$\,\buildrel >\over\sim\,$}}

\received{\today}
\revised{\today}
\accepted{\today}
\submitjournal{AJ}

\usepackage{graphicx}
\usepackage{float}

\shorttitle{HMXB Emission Deficit in NGC 7552}
\shortauthors{West et al.}
\graphicspath{{./}{figures/}}

\begin{document}

\title{The Large Deficit of HMXB Emission from Luminous Infrared Galaxies: \\
the Case of the Circumnuclear Starburst Ring in NGC 7552}

\correspondingauthor{Lacey West}
\email{lad012@uark.edu}

\author[0000-0002-5523-4723]{Lacey West}
\affiliation{Department of Physics, University of Arkansas, 226 Physics Building, 825 West Dickson Street, Fayetteville, AR 72701, USA}

\author[0000-0002-9202-8689]{Kristen Garofali}
\affiliation{NASA Goddard Space Flight Center, Code 662, Greenbelt, MD 20771, USA}

\author[0000-0003-2192-3296]{Bret D. Lehmer}
\affiliation{Department of Physics, University of Arkansas, 226 Physics Building, 825 West Dickson Street, Fayetteville, AR 72701, USA}

\author[0000-0003-3484-0326]{Andrea Prestwich}
\affiliation{Center for Astrophysics, Harvard-Smithsonian, 60 Garden Street, Cambridge, MA 02138, USA}

\author[0000-0002-2987-1796]{Rafael Eufrasio}
\affiliation{Department of Physics, University of Arkansas, 226 Physics Building, 825 West Dickson Street, Fayetteville, AR 72701, USA}

\author[0000-0002-4516-6042]{Wasutep Luangtip}
\affiliation{Department of Physics, Faculty of Science, Srinakharinwirot University, Bangkok 10110, Thailand}
\affiliation{National Astronomical Research Institute of Thailand, Chiang Mai 50180, Thailand}

\author[0000-0001-8252-6337]{Timothy P. Roberts}
\affiliation{Centre for Extragalactic Astronomy \& Department of Physics, University of Durham, South Road, Durham DH1 3LE, UK}

\author[0000-0001-8952-676X]{Andreas Zezas}
\affiliation{Center for Astrophysics, Harvard-Smithsonian, 60 Garden Street, Cambridge, MA 02138, USA}
\affiliation{Department of Physics, University of Crete, Heraklion GR-70013, Greece}
\affiliation{Institute of Astrophysics, FORTH, Heraklion GR-71110, Greece}

\begin{abstract}

Luminous infrared galaxies (LIRGs), the most extreme star-forming galaxies in the nearby (D~$<$~30~Mpc) Universe, show a notable X-ray emission deficiency (up to a factor of $\sim$~10) compared with predictions from scaling relations of galaxy-wide high mass X-ray binary (HMXB) luminosity with star-formation rate. In the nearby ($\approx$20~Mpc) LIRG NGC 7552, the majority of the IR emission originates in a circumnuclear starburst ring, which has been resolved into several discrete knots of star formation. We present results from recent Chandra observations of NGC 7552, which reveal significant deficits in the 2--7~keV X-ray luminosities from two of the most powerful star-forming knots. We hypothesize that the expected luminous HMXB populations in these knots are either (1) obscured by very large column densities or (2) suppressed due to the knots having relatively high metallicity and/or very young ages ($\lesssim$~5~Myr). We distinguish between these possibilities using data from recent NuSTAR observations, whose sensitivity above 10~keV is capable of uncovering heavily obscured HMXB populations, since emission at these energies is more immune to absorption effects. We find no evidence of a heavily obscured HMXB population in the central region of NGC 7552, suggesting suppressed HMXB formation. We further show that metallicity-dependent scaling relations cannot fully account for the observed deficit from the most powerful star-forming knots or the central region as a whole. Thus, we suggest that recent bursts in local star formation activity likely drive the high $L_{\rm{IR}}$ within these regions on timescales $\lesssim$~5~Myr, shorter than the timescale required for the formation of HMXBs.

\end{abstract}


\section{Introduction} \label{sec:intro}

The high mass X-ray binary (HMXB) X-ray luminosity function has been shown by several studies to scale with star-formation rate \citep[SFR; e.g.,][]{Mineo2012a, Lehmer2015, Lehmer2019, Luangtip2015}. However, \chandra\ observations of luminous infrared galaxies (LIRGs; $L_{\rm IR} > 10^{11}  L_\odot$), the most extreme star-forming galaxies in the nearby Universe, reveal a potential gap in our knowledge of HMXB formation. The X-ray emission from LIRGs can be significantly below the scaling-relation predictions \citep[up to a factor of $\sim$ 10; e.g.,][]{Iwasawa2009, Lehmer2010, Luangtip2015}. With distances typically $\gtrsim$ 50~Mpc, the X-ray binary populations in LIRGs are not often resolved by \chandra, making it difficult to identify the reason for the apparent X-ray emission deficiency in these galaxies.

Binary population synthesis models \citep[e.g.,][]{Linden2010, Fragos2013a, Belczynski2016, Kruckow2018, Artale2018, Wiktorowicz2017, Wiktorowicz2019} suggest that X-ray emission from HMXB populations is highly sensitive to local environmental properties, such as metallicity and star formation history (SFH). For example, \citet{Linden2010} estimate that the peak luminosity for the most luminous HMXBs in a population occurs $\sim$ 5--15 Myr after a star formation event and fades rapidly thereafter. This theoretical picture of HMXB formation is expected to be metallicity dependent, as metallicity influences the strength of stellar winds, and therefore stellar structure and evolution. At relatively low metallicities line-driven winds are weaker, resulting in HMXBs with tighter orbits and more massive stellar remnants. Lower metallicity environments can thus be expected to yield more numerous and luminous HMXB populations. At relatively high metallicity ($\gtrsim 1 Z_{\odot}$), the brightest systems are expected to be wind-fed HMXBs with supergiant donors, with production efficiency peaking $\sim$ 5 Myr post-starburst and fading very rapidly ($\sim$ few Myr), while at much lower metallicities HMXBs from the Roche lobe overflow formation channel lead to a secondary or prolonged formation peak extending to $\gtrsim$ 10 Myr. These effects lead to strong variations in the predicted HMXB population emission with age and metallicity, and recent empirical studies have shown evidence of these variations \citep[e.g.,][]{Antoniou2016, Brorby2016, Garofali2018, Antoniou2019, Fornasini2019, Fornasini2020, Lehmer2021, Gilbertson2022, Anastasopoulou2016, Anastasopoulou2019, Kouroumpatzakis2021}.

Given the above predictions, we suggest the following possibilities to explain the X-ray emission deficiency from LIRGs: (1) HMXBs are present, but severely obscured in the relatively low-energy ($\lesssim 7$~keV) X-ray band sampled by Chandra; (2) the formation of luminous HMXBs is suppressed by the high metallicity in these regions; and/or (3) the high infrared luminosities are due to very young ($\lesssim$~5~Myr) powerful bursts of recent star formation (rather than the relatively steady star-formation in more typical galaxies), and HMXBs have not yet had sufficient time to form. 

\begin{figure*}[t!]
  \centerline{
  \includegraphics[width=18cm]{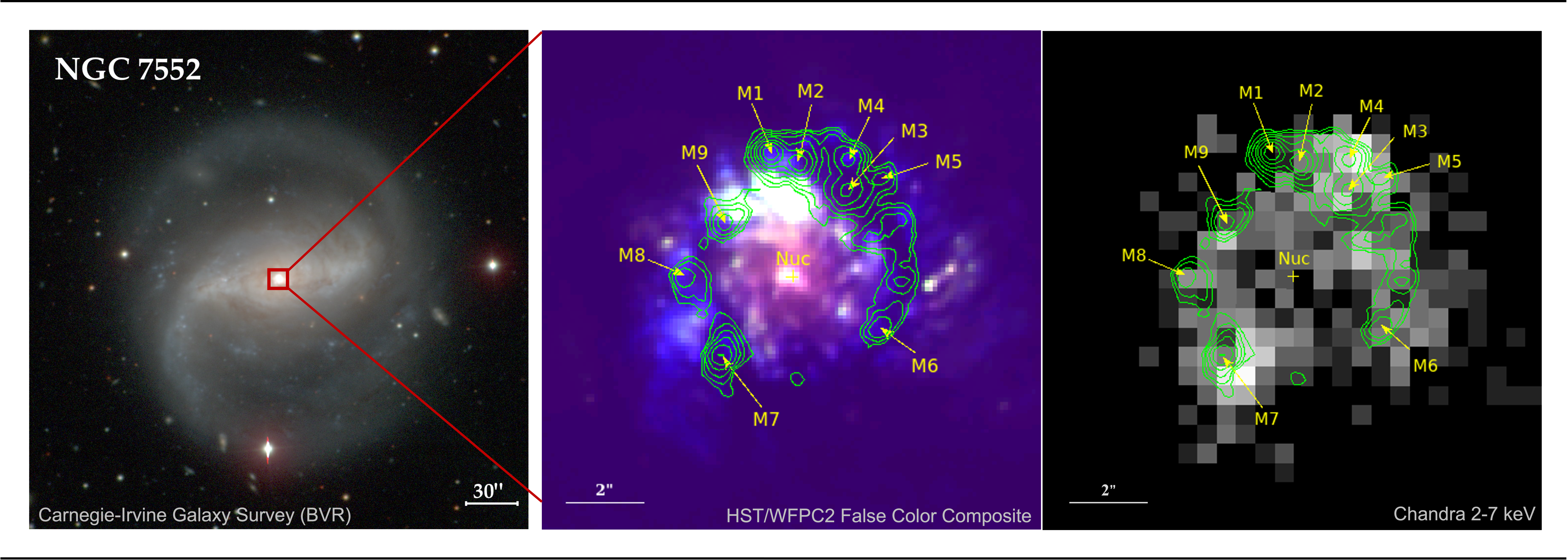}
  }
  \vspace{0.1in}
\caption{
({\it Left\/}) Carnegie-Irvine $BVR$ image of NGC 7552. 
({\it Center\/}) ``Zoomed in" false color composite image of the nuclear region of NGC 7552, with archival HST/WFPC2 images in the F439W (red), F336W (green) and F658N(H-$\alpha$, blue) filters and [Ne II] 12.8$\mu$m emission contours overlaid (green contours).  The locations of the nine prominent [Ne II] knots and $K$-band nucleus locations are indicated (yellow symbols and annotations). 
({\it Right\/}) \chandra\  2--7~keV image of the same region shown in the central panel.  The \xray\ emission appears to track the ring well with two potential point sources near the locations of M4 (RX-1) and M7 (RX-2).  We discuss these sources and their properties in more detail in Sections~\ref{sec:chandra} and \ref{sec:diff+ps}.
}
\label{fig:multwavimg}
\end{figure*}

Here we present a detailed investigation of NGC 7552 (see Figure~\ref{fig:multwavimg}), a relatively nearby ($\sim20$~Mpc) LIRG with an X-ray luminosity $\approx5$ times lower than expected from scaling relations \citep[][]{Luangtip2015}. NGC 7552 has been shown to not contain an active galactic nucleus \citep[AGN; e.g.,][]{ClaSah1992}; instead, the majority of the IR emission from this nearly face-on barred spiral galaxy comes from a circumnuclear starburst ring of diameter $\approx5$\arcsec\ ($\approx500$ pc), which has been resolved into several discrete clumps, or ``knots,'' of powerful star formation ($\sim$ 0.2--1.0 $M_{\odot}$~yr$^{-1}$ per knot) that spatially track the ring \citep[e.g.,][Figure \ref{fig:multwavimg}]{Forbes1994, Brandl2012}. From extinction-corrected mid-IR spectral lines, the ages of the knots are estimated to be young ($\approx$6 Myr) and highly obscured ($A_{V}$ $\approx 7$~mag). The central region is also reported to have a metallicity of $\sim1.5$ $Z_\odot$ \citep[][]{Moustakas2010, Wood2015}, which is relatively high compared to typical local galaxies. 

In this paper, we build upon results from previous multiwavelength studies of NGC 7552 and leverage the high spatial resolution of \chandra\ and hard-band sensitivity of \nustar\ to conduct an in-depth investigation into the apparent deficit of \xray\ luminous sources in this galaxy. We recently conducted deep ($\sim200$ ks entire) \chandra\ observations of NGC 7552 (PI: A. Prestwich), to study the 0.5--8~keV spectrum and resolve the circumnuclear starburst ring, and a $\sim200$ ks \nustar\ observation (PI: B. Lehmer), to probe the 3--30~keV emission and constrain the presence of any heavily obscured HMXB population. 

This paper is organized as follows: in Section \ref{sec:data}, we present the data sets utilized in this analysis and outline our X-ray data reduction procedures. In Section \ref{sec:methods}, we describe the specific spectral fitting technique used to model the 0.3--30~keV spectrum of the nuclear starburst ring from these X-ray data sets. In Section \ref{sec:results}, we provide the results of our X-ray spectral analyses. In Section \ref{sec:discussion}, we discuss and interpret these results in context with existing literature and discuss future directions for this work. Finally, we summarize our results in Section \ref{sec:summary}.

Throughout this paper, we calculate properties intrinsic to NGC~7552 (e.g., luminosities and SFRs).  These calculations assume a distance to NGC~7552 of 19.5~Mpc \citep{Tul1988} and make use of a \citet{kroupa2001} initial mass function (IMF).  Fluxes and luminosities are corrected for Galactic absorption, assuming a column density of $N_{\rm H, gal} = 1.0 \times 10^{20}$~cm$^{-2}$; however, these quantities are not corrected for intrinsic absorption within NGC~7552.

\section{Observations \& Data Reduction} \label{sec:data}

%
\begin{deluxetable*}{ccccc}
\centering
\tablewidth{\textwidth}
 \tablecaption{\chandra\ and \nustar\ Observations Used in this Work \label{tab:obs}}
\tablehead{
 \colhead{Obs. Start Date} & \colhead{Obs. ID} & \colhead{Inst.} & \colhead{Eff. Exposure (ks)} & \colhead{PI} \\
  \colhead{(1)} & \colhead{(2)} & \colhead{(3)} & \colhead{(4)} & \colhead{(5)} }

\startdata 
\hline 
& & {\chandra} & & \\
\hline 
2018-08-21 & 20267 & ACIS-S & 54.27  & A. Prestwich \\
2018-08-20 & 20268 & ACIS-S & 11.89  & A. Prestwich \\ 
2018-08-24 & 21675 & ACIS-S & 64.22  & A. Prestwich \\ 
2018-08-24 & 21676 & ACIS-S & 65.49  & A. Prestwich \\ 
\hline 
{\bf Total} & & & 195.87 &\\
\hline
& & { \nustar} & & \\
\hline 
2019-10-10 & 50501001002 & FPMA & 190.96 & B. Lehmer \\
2019-10-10 & 50501001002 & FPMB & 189.76 & B. Lehmer \\
\enddata 
\tablecomments{Col. (1): observation start date. Col. (2): observation ID. Col. (3): instrument. Col. (4): good time interval effective exposure times in ks after removing flared intervals. Col. (5): observation PI.}
\label{tab:obs}
\end{deluxetable*}
%

In this section, we describe relevant constraints from previous multi-wavelength studies of the nuclear region of NGC 7552, present the new \chandra\ and \nustar\ observations (listed in Table \ref{tab:obs}) employed in this paper, and describe our data reduction, astrometric alignment, and spectral extraction procedures.

\subsection{Multi-wavelength Data and Frame Alignment}\label{sec:multiwave}

\citet{Brandl2012} used high resolution mid-infrared spectroscopy from the VISIR and SINFONI instruments on the VLT, and Spitzer IRS, to create an IR spectral map of the central ring in NGC~7552 (see Figure 3a in Brandl et al. 2012) and constrain its properties. Nine unresolved mid-infrared (MIR) peaks (hereafter ``knots'' M1--M9) were identified (positions labeled in Figure \ref{fig:multwavimg}), tracing the central star-forming ring, seen also as a dark dusty ring in HST observations. The center panel of Figure~\ref{fig:multwavimg} shows three-band HST/WFPC2 imaging with [Ne~II] MIR contours highlighting the MIR peaks. The majority ($\approx$62\%) of the IR emission (and SFR) from all nine knots come from the three most intense components: M1, M2 and M7.  

\begin{figure*}[t!]   
\centerline{
\includegraphics[width = 8.5cm]{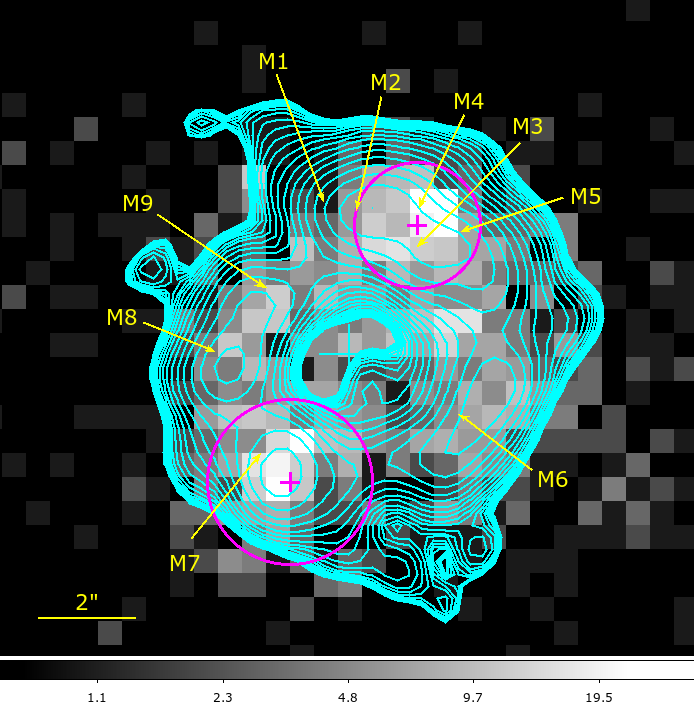}
\hfill
\includegraphics[width = 8.5cm]{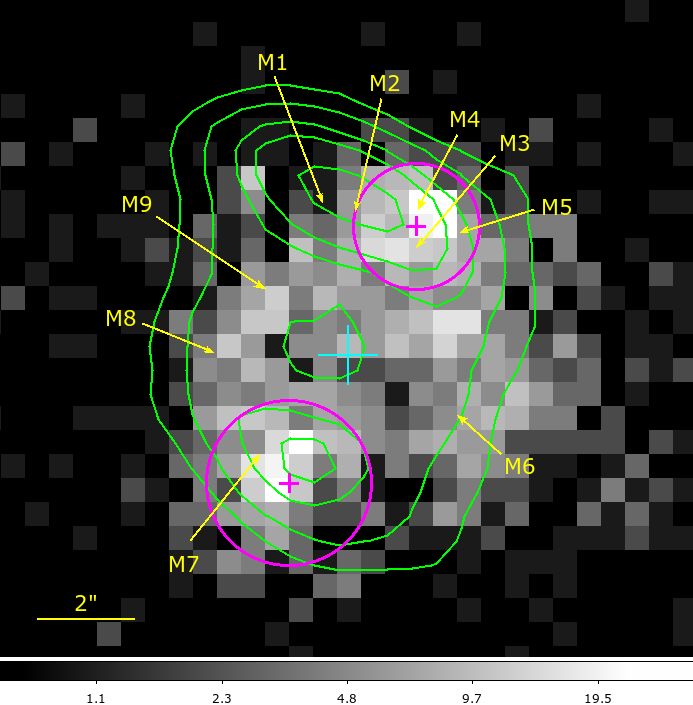}
}
\caption{\chandra\ 2--7~keV image (grayscale) of the nuclear region in NGC 7552 with \chandra-detected source locations (magenta circles with crosses at center), M1--M9 MIR peak positions (yellow labels), and ({\it left}) 3~cm radio continuum contours (cyan) and ($right$) molecular HCO+ contours (green) overlaid.  We find good alignment between the X-ray detected point sources RX-1 and RX-2 and radio continuum peaks, which are expected to be sites of recent supernovae and luminous HMXBs that emerge on timescales $\simgt$5~Myr following a star-formation event.  The peak molecular emission, however, appears to be offset from the X-ray sources, with the brightest region in the north also being associated with the locations of the brightest [Ne II] knots M1 and M2.  Given their association with strong molecular concentrations, M1 and M2 are expected to be younger than the knots associated with continuum radio and X-ray emission. 
}
\label{fig:contours}
\vspace{0.1in}
\end{figure*} 

SFR estimates, derived from the empirical $L_{\rm IR}$ scaling relation from \citet{Kennicutt1998}, for the nine individual knots range from $\sim$~0.2--1.0~$M_{\odot}$~yr$^{-1}$ (listed in Table \ref{tab:knot_props}), with a cumulative total of $\approx$~3.7~$M_{\odot}$~yr$^{-1}$. This sum of estimated individual knot SFRs is relatively low compared with previous estimates of the SFR in the entire nuclear region of NGC 7552 \citep[$\sim$~10--15~M$_\odot$~yr$^{-1}$; summarized in ][]{Pan2013}, but agrees well with the the nuclear region SFR value (6.1~M$_{\odot}$~yr$^{-1}$) extracted from the \citet{Lehmer2019} SFR map for NGC 7552 . Using extinction-corrected mid-IR spectral lines, the ages of the knots are estimated to be young ($\approx$5.6--6.3~Myr, with a small age range of $\approx$~0.8~Myr; listed in Table~\ref{tab:knot_props}), and highly obscured ($A_{V}~\approx$~5--9~mag).  Age estimates for M1--M9 were derived from equivalent widths (EWs) of extinction-corrected Br-$\gamma$ hydrogen recombination lines, assuming instantaneous star formation, and line-of-sight extinction was calculated from Br$\gamma$/Pa$\beta$ emission line ratios.  As discussed by \citet{Brandl2012}, {\it relative} ages estimated by this method are considered robust, however, absolute age estimates are more uncertain due to the impact of light from older generations of stars that may lie within the apertures.  The central region is also reported to have a relatively high metallicity \citep[$\sim$~1.5~$Z_\odot$;][]{Moustakas2010, Wood2015}, compared to typical local galaxies. 

Proper alignment between \chandra\ and the multiwavelength data is of great importance to this study. We chose to align all multiwavelength data sets to the \chandra\ frame using the following methodology.  The positions of the central point source and nine star-forming knots in the circumnuclear ring, which constitute the VLT frame, were provided by \citet{Brandl2012}.  \citet{Brandl2012} derived the absolute astrometry of the VLT frame (and quoted positions) by aligning the SINFONI $K$-band position of the nucleus to the 3~cm radio nucleus position from \citet{Forbes1994}.  To align the VLT/radio positions to the \chandra\ frame, we identified two 2--7~keV point-like sources, detected by \chandra\ (see $\S$\ref{sec:chandra} and the left panel of Figure~\ref{fig:contours}), within the circumnuclear ring that aligned well with two peaks in the 3~cm radio maps once shifted by $\approx$1.24\arcsec.  Throughout the remainder of this paper, we make comparisons between circumnuclear ring features and \xray\ measurements after applying this 1.24\arcsec\ shift to the VLT/radio source positions to bring them into alignment with \chandra.
  
After applying this astrometric shift to the VLT/radio data, we aligned HST images to the shifted radio frame (and thus \chandra\ frame) by applying an $\approx$0.8\arcsec\ shift to the HST frame by visually aligning the position of the central nuclear source in HST to the radio nucleus.  In the center panel of Figure \ref{fig:multwavimg}, we show the resulting HST imaging with the [Ne II] contours (from VLT VISIR) after performing our alignment procedure.  We find excellent agreement between the adjusted [Ne II] contours and the dust lanes in HST images.  The right panel of Figure \ref{fig:multwavimg} shows the 2--7~keV \chandra\ image with [Ne II] contours overlaid.  The overall ring pattern of the [Ne II] contours appears to match well the pattern observed with \chandra; however, it is clear from this imaging that there is strong variation in \xray\ brightness between [Ne II] peaks.  We discuss this further in Section~\ref{sec:discussion} below. 

\begin{figure*}[t!] 
    \centerline{
    \hspace{-75pt}
    \includegraphics[width=12cm]{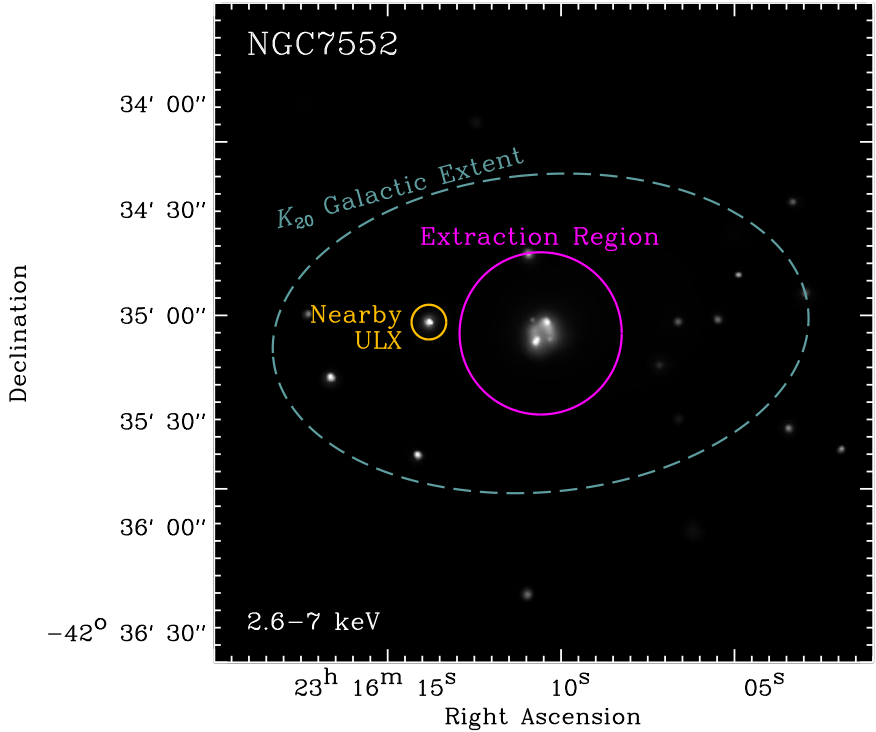}}
    \vspace{-5pt}
    \caption{  
    \chandra\ 2.6--7~keV image of NGC 7552. Spectral extraction aperture for the entire nuclear region is indicated in magenta; this aperture avoids contamination from the nearby ULX, J231613.6$-$423502.2, labeled in yellow.
}
\label{fig:chandra_n7552}
\end{figure*} 

\subsection{Chandra Data Reduction} \label{sec:chandra}

We obtained four \chandra\ observations of NGC 7552 for a cumulative 196~ks exposure from 2018 Aug 20 to 2018 Aug 24 (PI: A. Prestwich; see Table~\ref{tab:obs}). \chandra\ data reduction was carried out using CIAO~v.~4.8 with {\ttfamily CALDB}~v.~4.7.1,\footnote{http://cxc.harvard.edu/ciao/} following the procedure described in $\S$3.2 of \citet{Lehmer2019}. Briefly, we (1) reprocessed pipeline products using the {\ttfamily chandra\_repro} script; (2) removed bad pixels and columns and filtered the events list to include only good time intervals without significant ($>$3 $\sigma$) flares above the background level; (3) constructed merged events lists and astrometric solutions using the {\ttfamily merge\_obs} script; and (4) created additional products, including images, exposure maps and exposure-weighted PSF maps with a 90\% enclosed-count fraction, appropriate for the 0.5--2~keV, 2--7~keV and 0.5--7~keV bands.

At energies greater than $\sim2$ keV, \chandra\ observations resolve the circumnuclear ring into two discrete sources of emission, which we hereafter refer to as ``RX-1'' (23:16:10.66, $-$42:35:02.67) and ``RX-2'' (23:16:10.89, $-$42:35:07.94), located near the astrometrically-aligned MIR peaks M4 and M7, respectively (see Figures \ref{fig:multwavimg} and \ref{fig:contours}). Using CIAO~v.~4.11 with {\ttfamily CALDB}~v.~4.8.3, we use the {\ttfamily specextract} script to extract spectra from these two \chandra-detected sources with aperture sizes chosen to enclose a 90\% count fraction for each source (radii of 1.3\arcsec\ and 1.7\arcsec\ for RX-1 and RX-2, respectively). At the positions of the remaining MIR peaks, we use CIAO Tools in DS9 to extract counts from apertures of radii 0.75\arcsec\ (enclosed count fraction $\sim$84\%) for the purpose of calculating upper limits on the X-ray emission at these knot positions.

\begin{figure*}[t!]
  \centerline{
  \includegraphics[width=18cm]{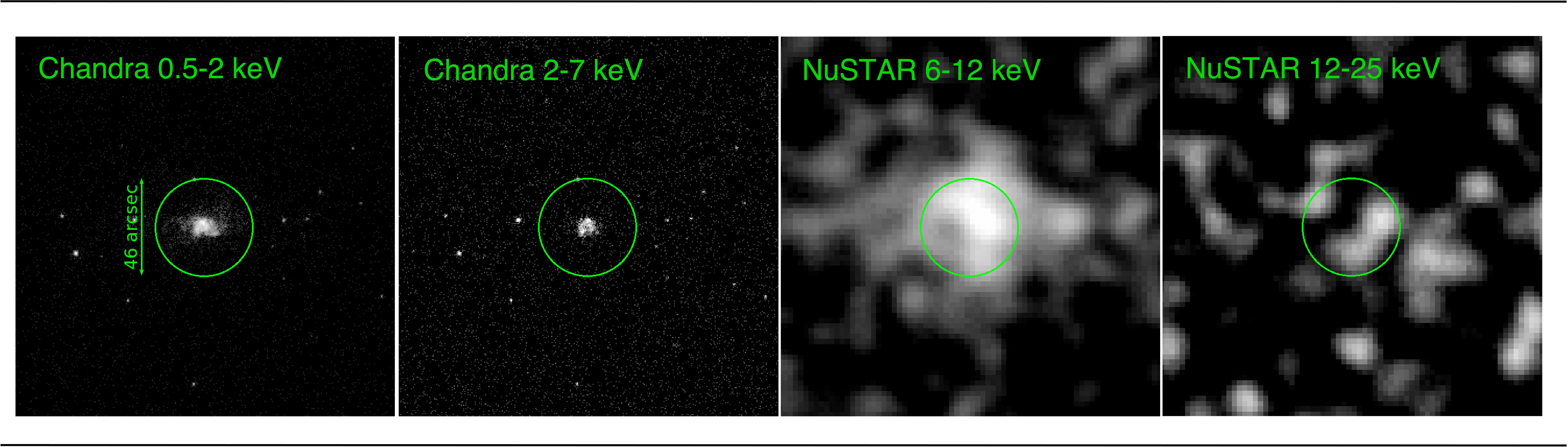}
  }
  \vspace{0.1in}
\caption{
Images of the NGC~7552 nuclear region from \chandra\ and \nustar.  From left-to-right, we show 0.5--2~keV, 2--7~keV, 6--12~keV, and 12--25~keV band images with the spectral extraction aperture for the entire nuclear region indicated as green circles with 46\arcsec\ diameters.  We find a very weak signal in the 12--25~keV band that is more consistent with a suppressed population of HMXBs, rather than a population that is simply buried behind a substantial obscuring column.
}
\label{fig:channu_img}
\end{figure*}

\chandra\ spectra for the {\it entire nuclear region} were extracted from a circular aperture of radius 23\arcsec, which corresponds to the extraction region that we use for our \nustar\ data (details provided in $\S$\ref{sec:nustar}).  In Figure \ref{fig:chandra_n7552}, we show this extraction region in magenta against the \chandra\ image.  We further extracted \chandra\ spectra from a {\it diffuse emission region}, which consisted of the entire nuclear extraction aperture with the two \chandra-detected point source apertures for RX-1 and RX-2 excluded (magenta regions in Figure \ref{fig:contours}); we use the diffuse spectrum to constrain local hot gas properties (e.g., temperatures and relative normalizations) and contributions from unresolved and undetected X-ray point sources, as described in $\S$\ref{sec:results}. \chandra\ background spectra were extracted from six large circular apertures (radii $\sim$15\arcsec) outside of the central nuclear region (as defined above), strategically placed to avoid contamination from prominent sources in the field. 

\subsection{NuSTAR Data Reduction} \label{sec:nustar}

We obtained a cumulative 191~ks \nustar\ exposure of NGC~7552 (ObsID 50501001002) over a single epoch from 2019 Oct 10 to 2019 Oct 14.  The \nustar\ data were analyzed using {\ttfamily HEASoft} v6.25 and \nustar\ Data Analysis Software ({\ttfamily NuSTARDAS}) v1.8.0 and CALDB v20181030.  Level~1 data were processed to level~2 using {\ttfamily nupipeline} under default settings.  This process filters out bad pixels, screens for cosmic rays, rejects high background levels (e.g., during passages through the SAA), and accurately projects events to sky coordinates, after accounting for dynamic relative offsets of the optical bench to the focal plane.

Given the relatively large PSF of \nustar\ ($\sim$60\arcsec\ HPD) versus \chandra, the central region of NGC~7552 is unresolved by \nustar\ and is effectively a point source (Figure \ref{fig:channu_img}).  We therefore utilize \nustar\ data to constrain the collective hard \xray\ emission of all components in the 23\arcsec\ radius nuclear region (defined above). Inspection of the \chandra\ data suggests that the majority of the 2--7~keV emission from NGC~7552 originates from the nuclear region; however, the relatively bright source J231613.6$-$423502.2 is located 33\arcsec\ from the center of the nuclear region and is $\approx$$1/3$ its brightness.  

Using {\ttfamily nuproducts}, we extracted the nuclear region and local background spectra for both FPMA and FPMB.  For the nuclear region, we utilized the 23\arcsec\ radius aperture ($\approx$33\% of the enclosed energy fraction for a point source), which contains minimal contributions from the nearby source J231613.6$-$423502.2, while maintaining relatively high S/N for the nuclear region data. The background spectra were extracted from three flanking circular apertures (each of radius 52\arcsec) that were chosen to lie on the same CCD chips in regions where the background rates were expected to be comparable to those in the source region \citep[i.e., after accounting for expected gradients; see, e.g.,][]{Wik2014}.

\section{Spectral Fitting Techniques} \label{sec:methods}

All spectral fits were performed using {\tt XSPEC} v12.10.0c \citep{Arnaud1996} on unbinned spectral data (i.e., in counts versus channel space) by minimizing the $C$ statistic \citep{Cash1979}.  Goodness-of-fit estimates for best-fit models were calculated following the procedure outlined in \citet{Bon2019}, which provides methods for calculating the expected $C$ statistic value $C_{\rm exp}$ and its variance $C_{\rm var}$ for a given Poisson model.  In the limit of large numbers of degrees of freedom, the $C$ distribution can be approximated as a Gaussian with mean $C_{\rm exp}$ and variance $C_{\rm var}$, and can be compared with measured values of $C$ to obtain the null hypothesis probability, which we define as:
\begin{equation}\label{eqn:pnull}
p_{\rm null} = 1 - {\rm erf}\left( \sqrt{\frac{(C - C_{\rm exp})^2}{2 \; C_{\rm
var}}} \right). 
\end{equation}
Errors on free model parameters were computed using the {\tt XSPEC error} command and reported as 90\% confidence intervals.

Rather than subtracting background, we develop background models to include as components in fits to unbinned source spectra. Briefly, we constrain the shape of the total background model for each instrument by fitting the background spectra extracted as described in $\S$\ref{sec:data}. \chandra\ (ACIS-S) and \nustar\ (FPMA/FPMB) total background models are developed following \citet{Bartalucci2014} and \citet{Wik2014}, respectively, with each consisting of both sky and instrumental background components \citep[e.g.,][]{Garofali2020}. We include the best-fit background models for each instrument in subsequent source fits as fixed components with background levels scaled according to the {\ttfamily BACKSCAL} keywords. 

For the purposes of this work, we are primarily interested in measuring the HMXB emission from SF knot regions in the NGC 7552 circumnuclear ring. Failure to properly account for the hot gas contribution to the flux in these regions can lead to elevated HMXB $L_{\rm X}$ estimates, even in harder bands (e.g., 2--7~keV). Each of our source spectral models includes separate components for 

\vspace{4in}
%
\begin{longrotatetable}
\begin{deluxetable*}{lccccccccccc}
\tabletypesize{\footnotesize}
\tablecaption{\chandra\ and \nustar\ Spectral Fit Results \\}
\tablehead{
%
\colhead{} & \colhead{} & \colhead{} & \colhead{} & \colhead{} & \colhead{} & \colhead{} & \colhead{} & \multicolumn{2}{c}{$\log(L_{{\rm X}})$} &    &  \\ 
%
\colhead{} & \colhead{} & \colhead{$C_{{kT}}$} & \colhead{$N_{{\rm H,gas}}$} & \colhead{$kT$}  & \colhead{$A_{{kT}}$}  & \colhead{$\Gamma$} &\colhead{$A_{{\Gamma}}$} & \colhead{(gas)} & \colhead{(HMXB)} & \colhead{}  & \colhead{} \\ 
%
\colhead{Source} & \colhead{Counts} & \colhead{(10$^{-2}$)} & \colhead{(10$^{22}$ cm$^{-2}$)} & \colhead{(keV)} & \colhead{(10$^{-5}$)} & \colhead{($\Gamma_1$,$E_{\rm b}$, $\Gamma_2$)} & \colhead{(10$^{-5}$)} & \multicolumn{2}{c}{(erg s$^{-1}$)} & \colhead{$C/N_{\rm chan}$} & \colhead{$p_{\rm null}$} \\
\colhead{(1)} & \colhead{(2)} & \colhead{(3)} & \colhead{(4)} & \colhead{(5)} & \colhead{(6)} & \colhead{(7)} & \colhead{(8)} & \colhead{(9)} & \colhead{(10)} & \colhead{(11)}  & \colhead{(12)} \\ } 

\startdata
\multicolumn{12}{c}{{\chandra}} \\
Diffuse$^a$ & 3614 &  \ldots & \ldots & 0.39$_{-0.02}^{+0.02}$ & 6.29$_{-0.32}^{+0.33}$ & \ldots & \ldots & \ldots & \ldots & \ldots & \ldots \\
\ldots  &  \ldots  &  \ldots & 0.62$_{-0.02}^{+0.02}$ & 0.84$_{-0.02}^{+0.02}$ & 30.8$_{-1.2}^{+1.3}$  & 1.8$^\star$ & 1.41$_{-0.16}^{+0.17}$ & 40.3 & 39.4 & 2083/2048 & 0.027 \\
RX-1$^b$ & 282 & 4.82$_{-0.45}^{+0.44}$ & $\dagger$ & $\dagger$ & $\dagger$  &2.02$_{-0.16}^{+0.16}$ & 0.83$_{-0.15}^{+0.17}$ & 39.0 & 39.1 & 1071/2048 & 0.273 \\
RX-2$^b$ & 284 & 4.17$_{-0.49}^{+0.49}$ & $\dagger$ & $\dagger$ & $\dagger$  &2.00$_{-0.13}^{+0.13}$ & 1.27$_{-0.18}^{+0.20}$ & 39.0 & 39.3 & 1266/2048 & 0.990 \\
\multicolumn{12}{c}{{Entire Nuclear Region Fit with \chandra\ + \nustar}} \\
Nuc (PL)$^c$ & 4168 & 106.34$_{-1.78}^{+1.74}$ & $\dagger$ & $\dagger$ & $\dagger$  & 2.10$_{-0.08}^{+0.08}$ & 4.52$_{-0.48}^{+0.53}$ & 40.4 & 39.8 & 3231/3398 & 0.392 \\
Nuc (BKNPL)$^d$ & 4168 & 108.71$_{-1.87}^{+1.86}$ & $\dagger$ & $\dagger$ & $\dagger$  & 1.88$_{-0.11}^{+0.11}$, 7.14$_{-0.55}^{+1.16}$, 3.33$_{-0.49}^{+0.76}$ & 3.51$_{-0.51}^{+0.56}$ & 40.4 & 39.8 & 3221/3398 & 0.192 \\
\enddata
\tablecomments{
Best-fit model parameters from spectral fits to (1) \chandra\ data for the diffuse region and point sources RX-1 and RX-2, and (2) \chandra\ plus \nustar\ data for the entire nuclear region. Quoted uncertainties are 90\% confidence intervals. Col.(1): Name of source region. Corresponding footnotes describe the spectral models employed in terms of {\tt XSPEC} models. Col.(2): Total number of counts used in spectral fit to the source region. Col.(3): Multiplicative constant modifying fixed diffuse model component as defined in the first two rows of the table. Col.(4): Column density in units of 10$^{22}$ cm$^{-2}$ for higher temperature {\tt APEC} component in the diffuse emission model. Col.(5): Plasma temperature in keV of {\tt APEC} model component(s). Col.(6): Normalization of the {\tt APEC} component(s). Col.(7): Photon index for power-law component, or indices and break energy in the broken power-law case. Col.(8): Normalization of the power-law/broken power-law component. Col.(9): 0.5--8 keV hot gas luminosity, measured using only the summed {\tt APEC} model components.  Col.(10): 0.5--8 keV HMXB luminosity, measured using only the power-law model/broken power-law components. Col.(11): Value of the $C$ statistic and number of spectral bins used in the fitting. Col.(12): Null hypothesis probability, as defined in Section~\ref{sec:methods}). \\}
\tablenotetext{\star}{Assumed value of the intrinsic photon index for the power-law model in the fit to the point-source-free spectrum (``diffuse'').  For all fits, the absorption of the power-law or broken power-law model component was fixed at $N_{\rm H, HMXB} = 3 \times 10^{21}$~cm$^{-2}$, the average value reported by \citet{Lehmer2019}.}
\tablenotetext{\dag}{
Parameters fixed to best-fit values from the fit to the point-source-free spectrum (``diffuse'') found in rows 1 and 2.}
\tablenotetext{{\tt a}}{
{\tt XSPEC} model: \texttt{tbabs$_{\rm Gal} \times$(apec$_{1} +$ tbabs$_{\rm gas}  \times$ apec$_{2} +$ tbabs$_{\rm HMXB} \times$ pow}). Thermal models assume $Z$ = 1.445 $Z_{\odot}$. Values of $kT_1$ and $A_{kT_1}$ are quoted in the first row (Col.(5) and (6)), while $kT_2$ and $A_{kT_2}$ are quoted in the second row in the same columns.}
\tablenotetext{{\tt b}}{
{\tt XSPEC} model: {\tt tbabs$_{\rm Gal} \times$(constant$_{kT} \times$(apec$_{1} +$ tbabs$_{\rm gas} \times$ apec$_{2}$)$+$ tbabs$_{\rm HMXB} \times$ pow}).}
\tablenotetext{{\tt c}}{
{\tt XSPEC} model: {\tt tbabs$_{\rm Gal} \times$(constant$_{kT} \times$(apec$_{1} +$ tbabs$_{\rm gas} \times$ apec$_{2}$)$+$ tbabs$_{\rm HMXB} \times$ pow}).}
\tablenotetext{{\tt d}}{
{\tt XSPEC} model: {\tt tbabs$_{\rm Gal} \times$(constant$_{kT} \times$(apec$_{1} +$ tbabs$_{\rm gas} \times$ apec$_{2}$)$+$ tbabs$_{\rm HMXB} \times$ bknpo}).}
\label{tab:spec_fits}
\end{deluxetable*}
\end{longrotatetable}
%
\vspace{1in}

\noindent the hot, diffuse gas and X-ray binary contributions (i.e., a power-law continuum), in addition to the fixed background components.  This allows us to calculate separately fluxes and luminosities for each source component. 

\section{Results} \label{sec:results}

To thoroughly investigate the apparent X-ray emission deficit in NGC 7552, we use newly-obtained \chandra\ and \nustar\ observations of this galaxy. In this section, we present results from spectral analysis of this X-ray data for each region of interest: point sources RX-1 and RX-2, the point-source-free diffuse emission, and the entire nuclear region. For each region we fit the unbinned source spectra using appropriate physical source models (as described below) and fixed, scaled background models (see Section~\ref{sec:methods}). Fit results are summarized in Table \ref{tab:spec_fits}.

\subsection{Point-Like Sources within the Nuclear Region} \label{sec:diff+ps}

In the 2--7~keV band, \chandra\ resolves the NGC 7552 nuclear region into two distinct point-like sources, RX-1 and RX-2 (see Figure \ref{fig:contours}), with additional unresolved emission tracing the star-forming ring. We attribute the unresolved X-ray emission primarily to hot, diffuse gas and unresolved HMXBs.

To properly account for hot gas emission in our analysis of the \chandra\ point sources and subsequent spectral fits, we first attempt to constrain temperature(s) and the flux contributions of the diffuse hot gas components using the point-source-free diffuse emission region as defined in Section~\ref{sec:methods} (i.e., 23\arcsec\ radius aperture with RX-1 and RX-2 excluded). Absorbed one or two temperature thermal plasma models are widely used to describe the hot interstellar medium in star-forming environments, particularly when fitting CCD resolution spectra \citep[e.g.,][]{Owen2009,Mineo2012b,Smith2018}. Therefore, we model the point-source-free diffuse emission region spectrum in {\tt XSPEC} using an absorbed two-temperature thermal plasma model ({\tt apec}) plus a power-law continuum (see Table~\ref{tab:spec_fits} notes for {\tt XSPEC} model details). We include both foreground Galactic and intrinsic absorption as convolution components ({\tt tbabs}) in our model and fix the abundance for the {\tt apec} components to 1.445~$Z_{\odot}$ \citep{Moustakas2010}. For the power-law component, we fixed the photon index to that of typical HMXBs ($\Gamma = 1.8$) to account for unresolved HMXB contribution. 

Free parameter best-fit values from fits to the diffuse emission region component with this spectral model are listed in the first two rows Table~\ref{tab:spec_fits}. The diffuse gas in NGC 7552 is well described by $\approx$ 0.4~keV and $\approx$ 0.8~keV plasmas, consistent with observations of the hot interstellar medium in other star-forming galaxies \citep[e.g.,][]{Grimes2005, Mineo2012b}. We include this diffuse model as a fixed component in all subsequent spectral fits, modified by a free multiplicative constant to allow for fluctuations in the local emission of hot gas within each spectral extraction aperture.

In the case of discrete \chandra\ point sources RX-1 and RX-2, we expect the majority of 2--7~keV emission to originate from HMXBs, which we model as simple absorbed power-laws. In addition to HMXB power-law component, we include the aforementioned diffuse hot gas component with fixed temperatures as constrained by the fit to point-source-free diffuse emission. We therefore allow three parameters to vary freely in each point source model fit: diffuse gas normalization (multiplicative constant), power-law photon index, and power-law normalization. As discussed in $\S$\ref{sec:methods}, we expect the power-law model component to effectively isolate emission from HMXBs. The best-fit parameters and associated uncertainties for each source are summarized in Table \ref{tab:spec_fits}, along with the 0.5--8~keV luminosities for the hot gas and HMXB model components.

The fit results for RX-1 and RX-2 are consistent with expectations for typical HMXB spectra ($\Gamma\approx$~2). With measured 0.5--8~keV luminosities for the HMXB component exceeding 10$^{39}$~erg~s$^{-1}$, we expect these sources are collections of HMXBs and/or ultra-luminous X-ray sources (ULXs). The proclivity toward steeper power-law indices may indicate RX-1 and RX-2 are indeed ULXs, which often show steepening in spectral slopes above $\sim$~3--5~keV \citep[e.g.,][]{Gladstone2009}; however, light curves from the three longest \chandra\ observations show a lack of flux variability between epochs for both RX-1 and RX-2, which may indicate  emission from these point-like objects is consistent with being produced by the collective emission of a population of underlying sources, rather than individual ULXs.

Given the relatively low S/N for each source from the \chandra\ data alone, the power-law indices are not strongly constrained, and we are unable to determine preference for a broken power-law model, which is typically used to describe the higher energy spectral curvature seen for ULXs.

\begin{figure}
    \centerline{
    \includegraphics[width=8.5cm]{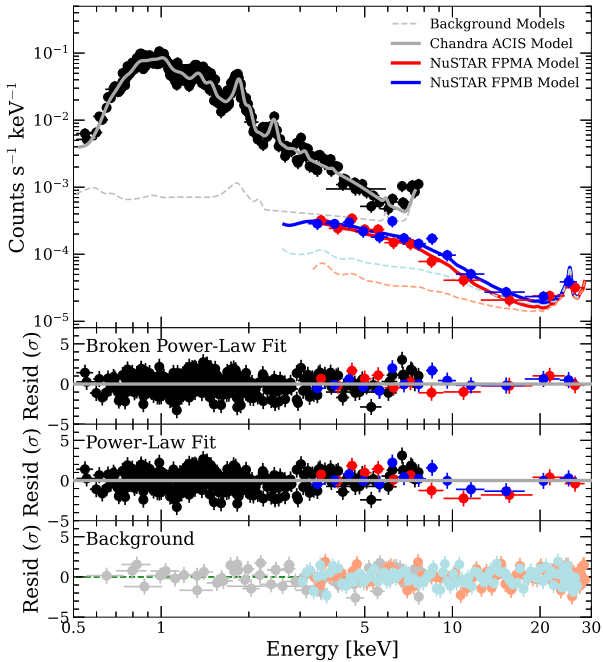}
    }
    \caption{Best-fit model to the combined \chandra\ and \nustar\ 0.5--30~keV spectrum of the entire nuclear region. Data are fit well by the combination of a two-temperature ($kT \sim$~0.4 and 0.8~keV) hot gas plus broken power-law model (solid curves), along with background model contributions (dashed curves).  The lower three panels show residuals for the on-source data for the broken power-law model (top residual row), the on-source data fit with a single power-law alternative to the broken power-law model (middle residual panel), and the background data and best-fit background model (bottom residual panel). The on-source data are best fit by a broken power-law with low-energy slope $\Gamma_{1} \approx$ 1.9, break at $E_{\rm b}\approx$ 7~keV, and high-energy slope $\Gamma_{2} \approx$ 3.3. Source and background models were fit to unbinned spectra; however, we have binned the spectra here for visualization purposes only. }
    \label{fig:spec}
\end{figure} 

\begin{figure}
    \centerline{
    \includegraphics[width=8.5cm]{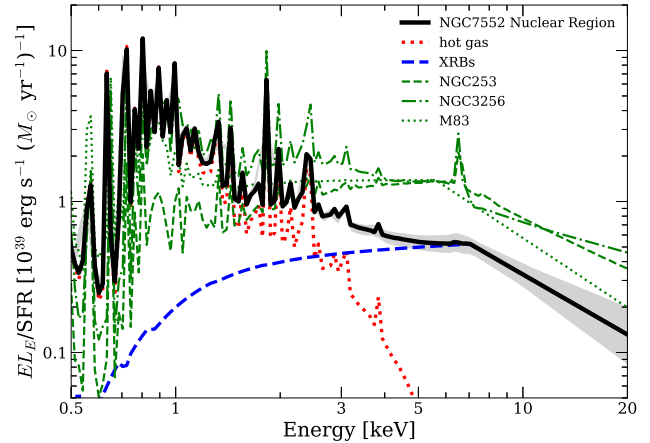}
    }
    \caption{Unfolded best-fit model spectrum, in terms of $EL_E$/SFR, of the entire nuclear region ({\it black curve\/}) and 1$\sigma$ model uncertainties ({\it grey shaded region\/}).  The hot gas and HMXB components of the model have been displayed as red-dotted and blue-dashed curves, respectively.  The HMXB fit is based on the broken power-law fit. For comparison, the equivalent model X-ray spectra are plotted for nearby star-forming galaxies NGC~253, NGC~3256, and M83, which have nearly solar metallicity.  Fits to these galaxies were based on the combination of \chandra/\xmm\ and \nustar\ observations that cover the full areal footprints of the galaxies \citep[see][for details]{Lehmer2015}. As discussed in $\S$\ref{sec:entire}, the estimated SFR from the central region of NGC 7552 is comparable to total galaxy-integrated SFRs from these comparison galaxies.
    }
    \label{fig:unf_spec}
\end{figure} 

\subsection{Entire Nuclear Region Spectral Analysis} \label{sec:entire}

When fitting the combined \chandra\ and \nustar\ spectrum for the entire nuclear region, we followed the same methodology that was applied to fitting the \chandra-detected point sources, and constructed a model consisting of hot gas and HMXB components. The parameters of the hot gas model component were again fixed to the values from the fit to the diffuse emission spectrum, i.e., those listed in the first two rows of Table~\ref{tab:spec_fits}, and we allowed only a multiplicative scaling factor, $C_{kT}$, to vary freely for this component. For the HMXB component, we considered both power-law and broken power-law models due to their success in describing the 0.3--30~keV spectra of star-forming galaxies \citep[see, e.g.,][]{Wik2014,Lehmer2015,Yukita2016,Garofali2020}. The results from our fits to the entire nuclear region using \chandra\ and \nustar\ data are listed in the last two rows of Table \ref{tab:spec_fits} and displayed in Figures~\ref{fig:spec} and \ref{fig:unf_spec}. 

In the top panel of Figure \ref{fig:spec}, we show the data for the entire nuclear region and our best-fit hot gas plus broken power-law model.  We further show residual panels, (data$-$model)/uncertainty, for best-fit models that include hot gas plus broken power-law, hot gas plus single power-law, and background models.  While formally, the fits that incorporate broken power-law and power-law are both statistically acceptable according to their $p_{\rm null}$ values (see Col.12 of Table \ref{tab:spec_fits}) when including both \chandra\ and \nustar\ data, we identified several negative residuals at 10--20~keV for our single power-law fits, suggesting that the power-law slope in this regime overestimates the data.  Indeed, the fit including the broken power-law allows for a steeper slope in this regime, and provides a significantly improved value of $C$ ($\Delta C = 10$).  In fact, when considering just the \nustar\ data at $E>10$~keV, the values of $p_{\rm null}$ are 0.075 and 0.829 for the single and broken power-law models, respectively.  This suggests that the broken power-law model indeed provides an improvement in the fits compared to the single power-law; hereafter, we consider the broken power-law to be our preferred model and refer to this model as our ``best model.''

The parameters for our best model indicate that the shape of the HMXB component (broken power-law with $\Gamma_{1}=1.88\pm0.11$, $E_{\rm b}=7.14^{+1.16}_{-0.55}$~keV, $\Gamma_{2}=3.33^{+0.76}_{-0.49}$) is similar to the spectral shape of HMXB populations in other well-studied star-forming galaxies with nearly solar metallicity (M83, NGC~253, and NGC~3256).  In Figure~\ref{fig:unf_spec}, we show our unfolded spectral model (in terms of $EL_E/$SFR) for the nuclear region of NGC~7552 across the 0.5--20~keV range, where our model is well constrained.  For comparison, we overlay the SED models of 
star-forming galaxies M83, NGC~253, and NGC~3256, which exhibit galaxy-integrated SFRs comparable to the total SFR of only the central region of NGC 7552.  It is clear from this representation that the HMXB component of our nuclear region best-fit model, which dominates above 3--5~keV, is suppressed (relative to other galaxies) systematically across the full spectral range.  Furthermore, we do not find any indications of a spectral ``upturn'' in the SED above 10~keV, which would have been expected for a buried HMXB population with luminosity consistent with local scaling relations.  This therefore suggests that the expected HMXB population is likely missing, instead of obscured (see Section~\ref{sec:discussion}).

Taking care to isolate the HMXB from the hot gas contribution, which can be substantial here, we calculate the flux (and therefore luminosity) of the HMXB component (power-law or broken power-law) of the source model. For the \chandra+\nustar\ fit to the entire nuclear region, we find the total $L_{\rm X}^{\rm HMXB} = 7.38^{+0.44}_{-0.43}\times10^{39}$ erg~s$^{-1}$ in 0.5--8~keV and ($3.59 \pm 0.20) \times10^{39}$ erg~s$^{-1}$ in 2--7~keV. This observed $L_{\rm X}^{\rm HMXB}$ is $\sim$3 times lower in 0.5--8~keV ($\sim$4 times lower in 2--7~keV) than predicted by $L_{\rm X}$/SFR scaling relations given the estimated SFR in this region (see $\S$\ref{sec:deficit}). 

\section{Discussion} \label{sec:discussion}

The nature, line-of-sight inclination, and proximity of NGC 7552 allow us to obtain critical insight into the physical processes leading to the X-ray deficit in this galaxy and potentially other, more distant LIRGs that have X-ray deficits up to a factor of $\sim$10 times below scaling-relation predictions. In this section, we use our new results, along with multiwavelength constraints from previous studies of NGC 7552, to quantify the $L_{\rm X}$/SFR constraints across the circumnuclear starburst ring and identify key drivers of the X-ray deficit. 

\begin{figure}[t!]
    \centering
    \includegraphics[width=9cm]{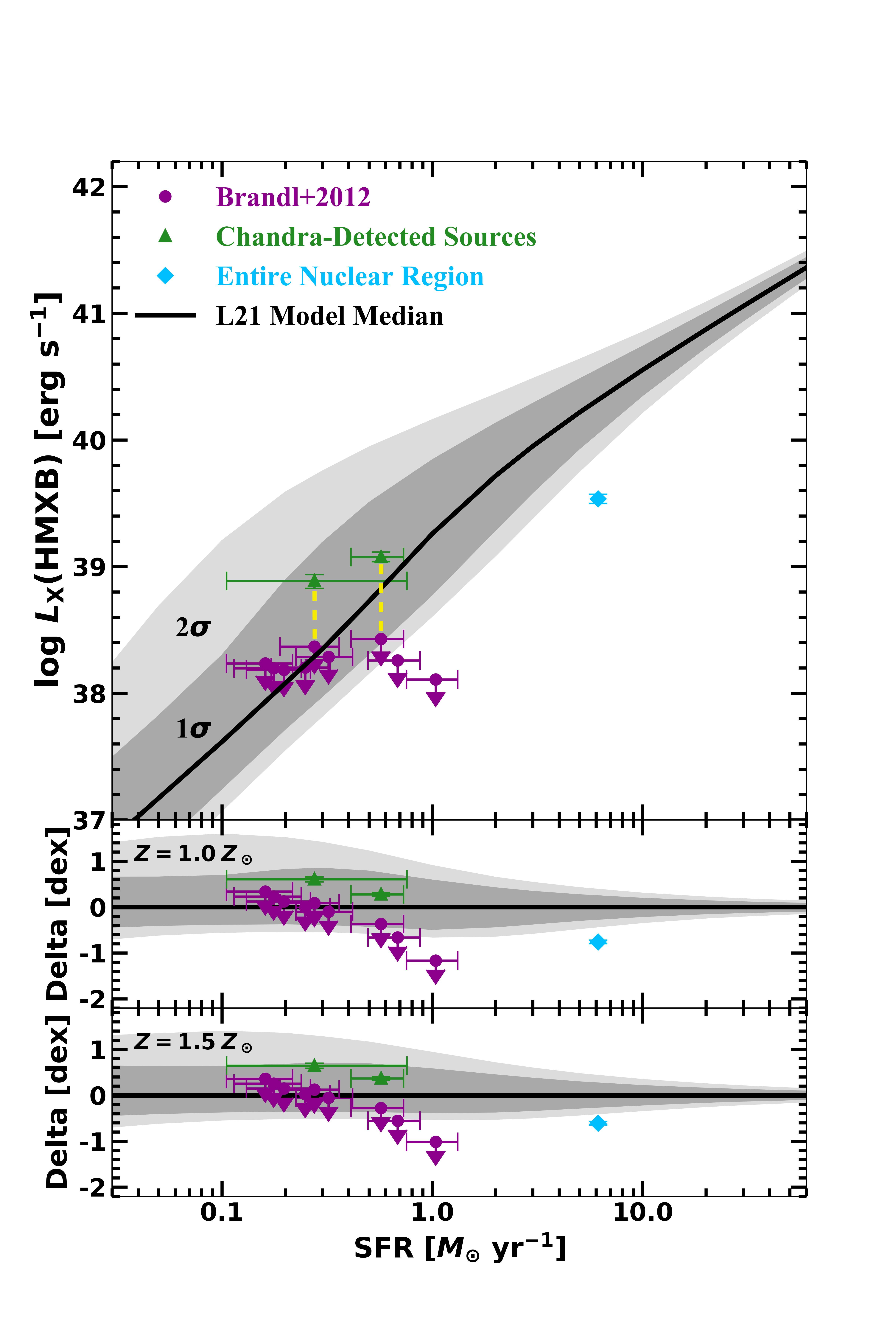}
    \caption{Observed HMXB X-ray luminosities (or upper limits) at 2--7~keV derived from \chandra\ at the location of each star-forming knot and \chandra-detected point source and from \chandra+\nustar\ for the entire nuclear region, along with SFR values at each location \citep{Brandl2012, Lehmer2019}. We compare $L_{\rm X}$--SFR measurements for each region with the $L_{\rm X}$(HMXB)--SFR relations for star-forming galaxies from \citet{Lehmer2021} at ({\it top two panels}) $Z\sim$~1~$Z_\odot$ and ({\it bottom panel}) $Z\sim$~1.5~$Z_\odot$.}
    \label{fig:L21relation}
\end{figure} 

\subsection{X-ray Luminosity/SFR Deficit for NGC 7552} \label{sec:deficit}

In Figure~\ref{fig:L21relation}, we present HMXB \xray\ luminosity and SFR constraints for the entire nuclear region and the locations of each of the nine star-forming knots (M1--M9).  SFR values for M1--M9 are as reported by \citet{Brandl2012}, which are based on local estimates of the total IR luminosity, converted to SFR using the \citet{Kennicutt1998} relation.  Since the majority of the knots do not have obvious coincident X-ray detections in \chandra, we calculated 3$\sigma$ upper limits on the X-ray luminosities at the knot locations.  These limits were calculated by converting 3$\sigma$ count-rate upper limits, 3$\sqrt{N_{\rm knot}}/t_{\rm exp}$ (where $N_{\rm knot}$ is the extracted 2--7~keV counts at each knot location and $t_{\rm exp}$ is the local vignetting-corrected exposure time), to flux and luminosity upper limits, using conversion factors based on the HMXB component from the ``best model" for the entire nuclear region. We present these limits in Figure~\ref{fig:L21relation} as magenta circles with downward-pointing arrows and report their local values in Table~\ref{tab:knot_props}.

Given their proximity to the circumnuclear ring, we expect the X-ray detections RX-1 and RX-2 are likely each associated with a star-forming knot.  However, neither detection is strictly coincident with a knot location, as shown in Fig. 1. For the discussion that follows, we present results assuming RX-1 and RX-2 are each associated with a single knot. This allows for associating a specific knot SFR with the X-ray luminosities measured for RX-1 and RX-2. In the case of RX-2, a unique association can be made to knot M7; however, for RX-1 there are a few knots crowded in the vicinity of the X-ray detection that may be potential matches (M3--M5), so we select association with M4, the knot with the smallest spatial offset from the position of the X-ray detection. We plot the X-ray luminosities for detections RX-1 and RX-2 as green triangles in Figure~\ref{fig:L21relation}, under the assumption these detections are associated with the SFRs measured for knots M4 and M7, respectively. However, given the small spatial offsets between RX-1 and RX-2 and these knots, we also present the associated X-ray upper limits at the positions strictly coincident with knots M4 and M7 in purple, connected to the detections for RX-1 and RX-2 via the dashed yellow lines. Given the proximity of RX-1 to knots M3, M4, and M5, the assignment of an estimated SFR from a single knot is less certain for RX-1 than for RX-2. Thus, we include a larger span of possible SFR values for RX-1, as indicated by the error bars in Figure 7, ranging from the lowest estimated knot SFR in the region (from knot M5) to the total combined SFR from all three nearby SF knots (M3, M4 and M5). This demonstrates that key interpretations are not affected by the choice of specific association of RX-1 with M4 and its SFR; the X-ray luminosity for the RX-1 detection remains consistent with the scaling relation within statistical scatter in all cases.

For comparison, we show the empirically-derived HMXB $L_{\rm X}^{\rm HMXB}$/SFR versus metallicity scaling relations from \citet[][hereafter, L21]{Lehmer2021} for $Z\sim$~1.0~$Z_\odot$ (top panel and middle residual) and $Z\sim$~1.5~$Z_\odot$ (bottom residual). L21 report a SFR and metallicity dependent scaling of the HMXB \xray\ luminosity function (XLF), which can be integrated to produce a total $L_{\rm X}^{\rm HMXB}$/SFR versus metallicity relation.  The median curves and uncertainty bands in Figure~\ref{fig:L21relation} appropriately account for the statistical variance expected for the total expected $L_{\rm X}$ by Monte Carlo sampling of the underlying HMXB XLF model.  The scaling relations are non-linear in the low-SFR regime due to poor statistical sampling of the full HMXB XLF when only a small number of luminous HMXBs are present.  More typically, {\it mean} $L_{\rm X}^{\rm HMXB}$--SFR scaling relations are used, which predict even higher $L_{\rm X}^{\rm HMXB}$ values in the low-SFR regime than those displayed in Figure~\ref{fig:L21relation}.

From Figure~\ref{fig:L21relation}, it is clear that the entire nuclear region and a few of the \xray\ upper limits at the star-forming knot locations are deficient at the  $> 2\sigma$ level for both relations (i.e., at 1 and 1.5~$Z_\odot$). The total observed 2--7~keV emission from the entire nuclear region is $\sim$4 times lower than predicted by the L21 relation given the SFR ($\approx$~6.1~M$_{\odot}$ yr$^{-1}$; adopted from SFR maps from Lehmer et al. 2019).

It is interesting to note that the \chandra-detected knots M4 (RX-1) and M7 (RX-2) and some upper limits (M3--M9) are consistent with L21 scalings at 1--1.5~$Z_{\odot}$, suggesting that some of the regions within the ring may be more like low-SFR galaxies (e.g., $\approx$5--100~Myr old populations harboring HMXBs). However, the entire nuclear region and the most highly active star-forming knots (M1 and M2) seem significantly deficient relative to these relations.  Although upper limits do not reflect tight constraints on X-ray emission levels, it is notable that, among the star-forming knots, the magnitude of the $L_{\rm X}^{\rm HMXB}$ deficit increases with increasing SFR. In the following section, we discuss some possible explanations for these X-ray emission deficits. 

%
\begin{deluxetable*}{lccccccc}
\tabletypesize{\small}
\tablewidth{\textwidth}
\tablecaption{Summary of M1--M9 Region Properties \label{tab:knot_props} \\}
\tablehead{
\colhead{} & \colhead{} & \colhead{} & \colhead{SFR} & \colhead{Age} & \colhead{Counts} & \colhead{Predicted Counts} & \colhead{log($L^{\rm HMXB}_{\rm X, lim}$)} \\
\colhead{Source} & \colhead{$\alpha$ (J2000)} & \colhead{$\delta$ (J2000)} & \colhead{(M$_{\odot}$ yr$^{-1}$)} & \colhead{(Myr)} & \colhead{(2--7~keV)} & \colhead{(2--7~keV)} & \colhead{(2--7~keV)}\\
\colhead{(1)} & \colhead{(2)} & \colhead{(3)} & \colhead{(4)} & \colhead{(5)} & \colhead{(6)} & \colhead{(7)} & \colhead{(8)} \\
} 
\startdata 
M1 & 23:16:10.82 & $-$42:35:02.5 & $1.04\pm0.28$ & 5.6 & 32 & 140--1226 & 38.1 \\
M2 & 23:16:10.75 & $-$42:35:02.7 & $0.68\pm0.19$ & 5.8 & 64 & 75--724 & 38.3 \\
M3 & 23:16:10.65 & $-$42:35:03.5 & $0.32\pm0.10$ & 5.9 & 73 & 24--257 & 38.3 \\
M4 & 23:16:10.64 & $-$42:35:02.7 & $0.27\pm0.09$ & 5.5 & 106 & 18--204 &  38.4 \\
M5 & 23:16:10.56 & $-$42:35:03.2 & $0.16\pm0.06$ & 5.8 & 58 & 7--89 &  38.2 \\
M6 & 23:16:10.57 & $-$42:35:07.0 & $0.18\pm0.06$ & 6.3 & 48 & 9--102 &  38.2 \\
M7 & 23:16:10.93 & $-$42:35:07.6 & $0.57\pm0.16$ & 6.2 & 138 & 56--568 &  38.4 \\
M8 & 23:16:11.02 & $-$42:35:05.7 & $0.20\pm0.07$ & 6.3 & 45 & 10--118 & 38.2 \\
M9 & 23:16:10.92 & $-$42:35:04.3 & $0.25\pm0.08$ & 6.3 & 49 & 14--165 &  38.2 \\
\enddata
\tablecomments{Properties of M1--M9 regions as reported in \citet{Brandl2012} and as determined from our X-ray analysis. Col. (1): MIR source ID, as assigned by \citet{Brandl2012}. Col. (2): astrometrically-aligned RA coordinates. Col. (3): astrometrically-aligned DEC coordinates. Col. (4): local SFR, derived from $L_{\rm IR}$ reported by \citet{Brandl2012}. Col. (5): MIR knot age estimate, as reported by \citet{Brandl2012}. 
Col. (6): Total raw 2--7~keV X-ray counts measured in region ($R = 0.75$\arcsec). Col. (7): Range of predicted raw 2--7~keV counts (16th--84th percentile) for each knot SFR. 
Col. (8): $L^{\rm HMXB}_{\rm X}$ upper limit, calculated as $3 \sqrt{N} /t_{\rm exp}$, aperture-corrected, and converted to HMXB flux using the nuclear region best-fit broken power-law model.}
\label{tab:knot_props}
\end{deluxetable*}

%

\subsection{What Drives the Deficit of X-ray Emission in NGC 7552?} \label{sec:whydef}

\subsubsection{The Impact of Obscuration} \label{sec:obscuration}

Previous studies have suggested obscuration as a leading factor in suppressing the observed 2--10~keV emission from LIRGs that fall below the $L_{\rm X}$/SFR correlation \citep[e.g.,][]{Lehmer2010, Luangtip2015}. Although the entire nuclear region appears as a single blended source of emission due to \nustar's relatively low spatial resolution (see Figure \ref{fig:channu_img}), we expect \nustar's sensitivity at $\gtrsim10$~keV to be sufficient for detecting the signatures of a heavily obscured HMXB population ($N_{\rm H} \simgt 10^{24}$~cm$^{-2}$), if present, in NGC~7552. 

In a scenario where a heavily obscured HMXB population does exist in the central region of NGC 7552, we expect an overall upturn in the X-ray SED at energies $E>10$~keV, which would be observed as a relatively shallow spectral slope from 8--30~keV.  However, as shown in Section~4.2, our best-fit models prefer a relatively steep power-law, or steepening broken power-law, in this portion of the spectrum that is consistent in shape with other star-forming galaxies that do not show a deficit in X-ray emission relative to the local $L_{\rm X}$/SFR correlation. 

Furthermore, as shown in Figure \ref{fig:unf_spec}, comparing the SFR-normalized spectrum of the nuclear region in NGC 7552 with the equivalent model for other nearby star-forming galaxies reveals similar hot gas emission levels with a clear divergence at energies $\gtrsim$3~keV, where the HMXB contribution dominates the overall emission from NGC 7552. To explain the observed spectral curvature for the NGC 7552 nuclear region while invoking a heavily buried HMXB population to explain the deficit in hard band emission (i.e., Figs 6--7) would require extreme levels of obscuration ($N_{\rm H}\gtrsim$~10$^{24-25}$~cm$^{-2}$). \citet{Brandl2012} utilized the VLT SINFONI Br$\gamma$/Pa$\beta$ ratio to infer levels of extinction to the star-forming regions in the ring, finding extinction levels of $A_V \approx$~5--10~mag.  Using the relation $N_{\rm H}/A_V \approx 2.21 \times 10^{21}$~cm$^{-2}$~mag$^{-1}$, derived from observations in the Milky Way \citep[e.g.,][]{Guver2009}, implies an $N_{\rm H}~\simlt 10^{22}$~cm$^{-2}$, far below the level of obscuration required to suppress the X-ray emission below that measured by our observations. Thus, the combination of the \nustar\ and mid-IR constraints strongly indicate that obscuration is unlikely to account for the observed X-ray emission deficit in NGC 7552. 

Given the lack of evidence for a luminous, heavily obscured X-ray emitting population in the central region of NGC 7552, we suggest that the expected luminous HMXB population is more likely not present due to metallicity and/or age effects. In the next two sections, we explore these possibilities in turn.

\subsubsection{Metallicity Dependence of HMXB Emission} \label{sec:metallicity}

Over the last decade, several studies have established that the $L_{\rm X}$/SFR ratio declines with increasing metallicity \citep[e.g.][]{Basu2013,Basu2016,Brorby2016,Fornasini2019,Fornasini2020,Lehmer2021}.  Such a decline has been predicted by HMXB population synthesis models, as a result of the effects of stellar winds on binary evolution.  High-metallicity systems lose more mass and angular momentum over binary lifetimes than low-metallicity systems, leading to wider binary separations, lower-mass compact-object remnants, and lower luminosity HMXB populations \citep[see, e.g.,][]{Linden2010,Fragos2013a,Fragos2013b,Wiktorowicz2017}.

Recent work by L21, which places constraints on the  metallicity dependence of the HMXB XLF and detailed $L_{\rm X}^{\rm HMXB}$--SFR relation, enables us to investigate the extent to which metallicity plays a role in the apparent suppression of X-ray emission from the central region of NGC 7552. In the bottom two panels of Figure~\ref{fig:L21relation}, we present data residuals for the L21 relations at $\approx$1~$Z_{\odot}$ and $\approx$1.5~$Z_{\odot}$, respectively. 

Due to the decline of $L_{\rm X}$/SFR with increasing metallicity, the $L_{\rm X}^{\rm HMXB}$--SFR relation associated with $\approx$1.5~$Z_{\odot}$ better describes data from NGC 7552; however, the entire nuclear region and most highly star forming knots (M1 and M2) are still significantly ($>$2$\sigma$) deficient in X-ray emission. Metallicity is not expected to vary drastically between the star-forming knots in this system, since the gas that fuels the SF in each knot is likely to have funneled in through the bar--ring ``contact points'' in a relatively consistent manner for all knots. Thus, our results suggest that metallicity alone is unable to account for the significant deficit in \xray\ emission in NGC~7552.  

\subsubsection{Stellar Age Effects} \label{sec:age}

In addition to the effects of metallicity, it is possible that the deficit in HMXB emission from the circumnuclear ring in NGC~7552 is due to the very short timescales of star-formation in the ring not being sufficiently long (i.e., $\simgt$5~Myr) to have formed the compact objects required to power HMXBs.  Predictions from population synthesis models \citep[e.g.,][]{Linden2010} suggest peak HMXB emission occurs on timescales $\sim$~5--15~Myr following a burst of star formation.  The deficit in HMXB emission from the entire nuclear region and the most highly star-forming knots suggests the absolute ages for at least some of the star-forming knots may actually be $<$ 5 Myr (i.e., too young for HMXBs to have formed), and that the ages of the remaining knots ($\sim$ 5--6 Myr) correspond to the timescale on which HMXB production is ``ramping up" following a recent burst of star formation.

As discussed in $\S$\ref{sec:multiwave}, 
ages for the individual star-forming knots have been estimated by \citet{Brandl2012} using EWs of the Br-$\gamma$ hydrogen recombination line assuming single simple stellar population models.  \citet{Brandl2012} validate reported absolute age estimates ($\sim$5.5--6.3~Myr) using two independent methods, and further constrain the age of knot M1 to $\gtrsim$4~Myr by comparing observed line flux ratios with the starburst models of \citet{Snijders2007}. However, \citet{Brandl2012} emphasize the largest potential source of error in the estimated ages may be the assumption that apertures contain a single coeval population. It is possible that each MIR peak contains multiple smaller clusters, as the formation of one massive cluster in a dense environment can trigger the formation of additional ``second-generation" clusters nearby \citep[e.g.,][]{Deharveng2003, Deharveng2005}, thereby skewing the original cluster age estimates to somewhat older light-weighted average value. 

Physical interpretations of galaxy bar--starburst ring interactions and dynamics within barred spiral galaxies, like NGC 7552, vary widely. In one proposed mechanism \citep[``pearls on a string",][]{Boker2008}, gas is thought to travel radially inward along bar dust lanes and accumulate at two contact points, which are approximately perpendicular to the position angle of the bar major axis \citep{Schinnerer1997,ReganTeuben2003}. Gas then travels into the central ring through these bar--ring contact points and ignites short-lived bursts of SF activity as gas density becomes sufficient.  Clusters then orbit along the ring, aging as they travel away from the contact points. 

\citet{Brandl2012} provide some evidence for an age gradient within the central region of NGC 7552, where the youngest clusters are positioned nearest to the contact points between the galaxy bar and starburst ring. This finding suggests that most of the gas enters the ring at these contact points and agrees with results from \citet{Mazzuca2008} indicating a positional correlation between the youngest H II regions and the contact points in the majority of barred galaxies \citep[see also, e.g., NGC 1068, NGC 7771, M100, IC 4933, M83; ][]{Davies1998,Smith1999,Ryder2001,Allard2005,Ryder2010,Knapen2010}. The influx of gas at these points might ignite bursts of star formation, and are incorporated into the ring, and continues to rotate relative to the galaxy bar. In such a scenario, the discontinuous supply of gas could result in very bursty star formation at the bar-ring contact points, driving high levels of IR emission (and thus SFR estimates) at these locations on timescales shorter than those required for HMXBs to form in abundance.  
 
\begin{figure}[t!]
\centerline{
\includegraphics[width=8cm]{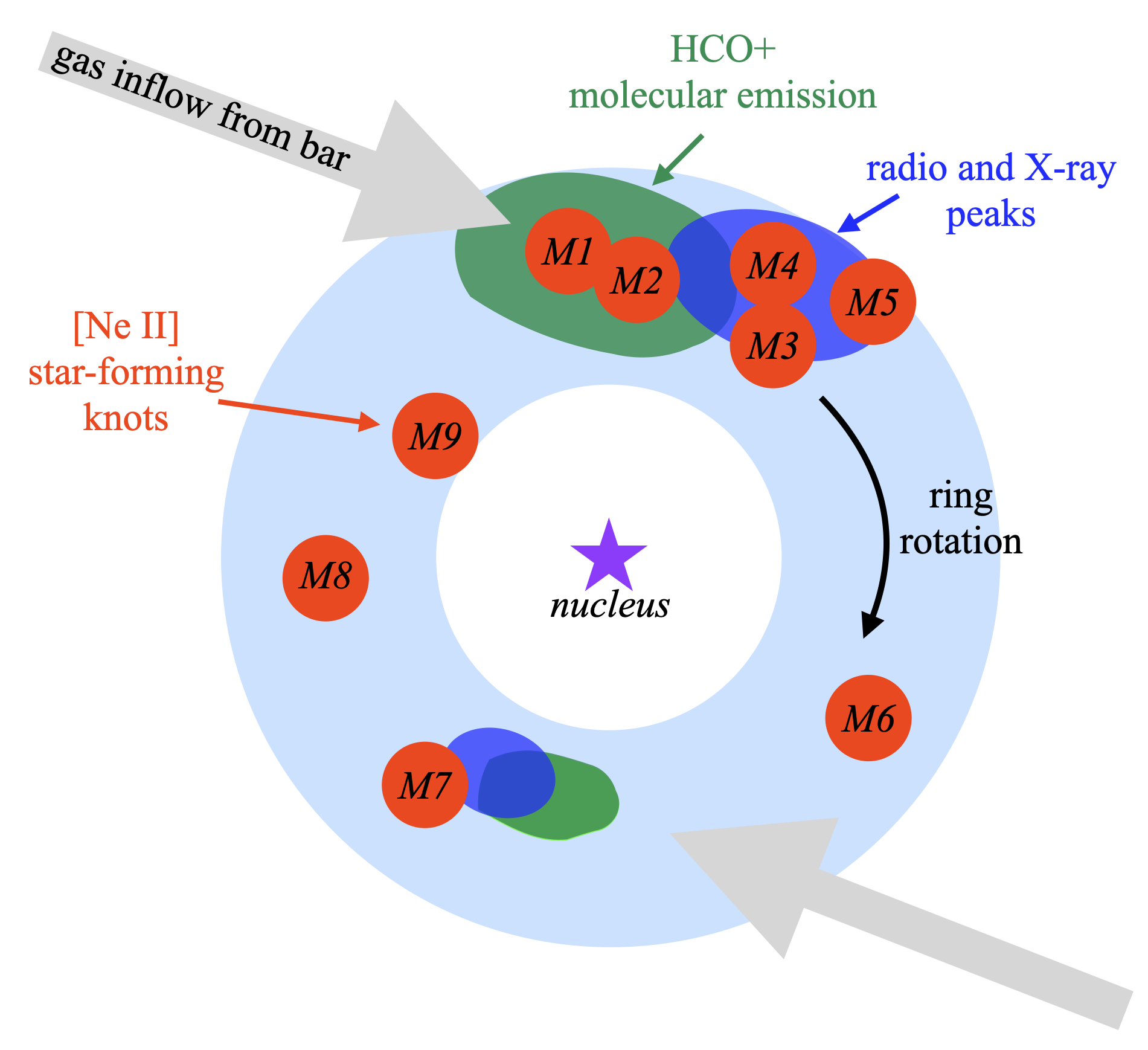}
}
\caption{
Cartoon depiction of our current model.  The circumnuclear ring (light blue) is fed by gas inflow from the bar (gray arrows) and accumulates at contact points in the north and south portions of the bar, where molecular concentration are observed as HCO+ emission (green regions).  As gas accumulates near the contact points, star-formation triggers and joins the rotating circumnuclear ring.  The youngest star-forming knots (red circles), M1 and M2, which are observed through [Ne II] line emission are close to the HCO+ overdensities and older intense knots of star-formation (e.g., M3--M5 and M7) are found to be associated with peak radio and X-ray emission (blue regions) due to the onset of supernovae and XRB populations, respectively.  See Figure~\ref{fig:contours} for data motivating this picture.
}
\label{fig:cartoon}
\end{figure}

For the case of NGC~7552, \citet{Pan2013} presented direct evidence of an age gradient between the contact point location near M1 and downstream ring locations (clockwise from M1) using the relative locations of molecular line and radio continuum peaks in the circumnuclear ring.  In Figure~\ref{fig:cartoon}, we illustrate this picture in cartoon form.  Specifically, molecular emission from HCO$^+$, which is indicative of very dense molecular clouds and triggered sites for star-formation, is observed to show two peaks on opposite sides of the ring where dust lanes from the bar in NGC~7552 meet the circumnuclear ring.  The radio continuum emission at 3~cm, which is indicative of synchrotron emission from supernovae, shows a similar peak pattern, but displaced clockwise away from the HCO$^+$.  
We find that the \chandra\ 2--7~keV \xray\ detections appear to be closely matched to the two prominent 3~cm peaks (see Fig.~\ref{fig:contours}), as would be expected in the scenario where HMXBs begin to ``turn on'' at timescales associated with core-collapse supernovae and the formation of the compact objects.  These lines of evidence are consistent with a scenario where the ages of the bright M1 and M2 star-forming knots, which are also coincident with the molecular peak from HCO$^+$ (see Fig.~\ref{fig:contours}), are too young ($\simlt$5~Myr) to have formed compact objects and HMXBs, leading to observed deficits in the $L_{\rm X}/$SFR ratio for these regions and the circumnuclear region as a whole.  Furthermore, the knots located downstream from the HCO$^+$ peaks that are coincident with the radio emission peaks (M3, M4, M5, and M7) may be somewhat older ($\simgt$5~Myr) and as such have X-ray properties consistent with $L_{\rm X}$/SFR correlations.

Given the above, we regard the stellar age effects on the formation of HMXBs to be the most likely explanation for the deficit of X-ray emission in NGC~7552. The observed X-ray deficit in the circumnuclear region of NGC~7552, coupled with the fact that the spectral {\it shape} for this region is consistent with the SEDs for other starburst galaxies albeit at a suppressed level suggests we may be probing a ``ramp-up" in HMXB production in this extreme star-forming ring, relative to the likely ``equilibrium" $L_{\rm X}^{\rm HMXB}$/SFR measured from galaxy-integrated populations in other nearby starbursts.
This result motivates future multiwavelength studies of other LIRGs with X-ray deficits.  In particular, studies that utilize high spatial resolution infrared data to study the properties of star-forming complexes in LIRGs \citep[e.g., the JWST GOALS program;][]{Evans2022, Inami2022, Lai2022} would help clarify the impact of recent star-formation history on X-ray emission in galaxies.   Perhaps, if physical processes occurring in these extreme star-forming regions can be understood in detail, age effects can be characterized on short timescales such that X-ray measurements could eventually serve as quantitative constraints on absolute ages.

\section{Summary} \label{sec:summary}

In this work, we have utilized $\sim$ 200~ks of both \chandra\ and \nustar\ observations to investigate the apparent X-ray emission deficit from the relatively nearby, nearly face-on LIRG NGC 7552 and build upon previous multiwavelength studies of this galaxy. 

By pairing the high spatial resolution of \chandra\ with the hard-band sensitivity of \nustar, we conducted an in-depth investigation into the apparent X-ray emission deficit in this galaxy, finding the deficit is likely driven by bursty star formation in the circumnuclear ring resulting in regions that are too young to have formed compact objects and HMXBs.
Our key findings can be summarized as follows:

\begin{itemize}
    \item[--] \chandra\ resolves the circumnuclear starburst ring of NGC 7552 into two 2--7~keV detected point-like sources, which have spectra that are modeled by steep, though not tightly constrained, power-law indices. These sources have observed 0.5--8~keV luminosities of $\approx$$10^{39}$~\lum, consistent with being collections of HMXBs and/or ULXs.
    
    \item[--] The combined \chandra\ and \nustar\ spectra for the entire nuclear region are well-described by a two-temperature thermal plasma dominating at $E \simlt 3$~keV and a broken power-law at higher energies, associated with the HMXB population, that steepens in slope at $E \simgt 7$~keV.  When compared with the HMXB spectra of other, more typical, star-forming galaxies in the local universe, the SFR-normalized spectrum of the entire nuclear region of NGC~7552 is deficient by a factor of $\approx$3--4 out to at least 20~keV.  Furthermore, we do not find evidence for any upturn in the spectrum at $E>10$~keV that would be expected if a luminous, buried HMXB population were present.

    \item[--] We tested the impact of metallicity on the HMXB population in NGC~7552, by comparing the $L_{\rm X}$/SFR measurement of the entire nuclear region with recent metallicity-dependent scaling relations.  We find that even at a supersolar metallicity (1.5~$Z_\odot$), the expected \xray\ luminosity of this region is deficient by $>$2$\sigma$, suggesting that metallicity alone cannot explain the observed \xray\ deficit. 

    \item[--] Using the \chandra\ data, we extracted 2--7~keV counts at the locations of nine infrared star-forming knots, as defined by \citet{Brandl2012}, and computed upper limits on 2--7~keV HMXB luminosity to compare with recent empirical $L_{\rm X}$--SFR relations from the literature. The most highly star-forming knots are clear outliers, with deficits from predicted emission levels seeming to worsen with increasing SFR.  However, the two \xray-detected sources within the circumnuclear star-forming ring appear to have properties consistent with the scaling relations. 

    \item[--] We find that the star-forming knots that were significantly deficient in \xray\ emission appear to be associated with concentrations of dense molecular gas traced by HCO$^+$, while the knots with \xray\ properties consistent with the $L_{\rm X}$--SFR relation are more closely associated with peak radio continuum emission at 3~cm.  The timescale for star-formation in the knots, as traced by the equivalent width of Br$\gamma$ line emission suggest very young ages of $\simlt$6--7~Myr, albeit with large uncertainty on absolute age estimates.  Our results suggest that the \xray\ deficient knots are likely younger than the non-deficient knots, as a result of the delay time required for HMXB formation following a burst of star formation.
    
    \item[--] We have provided useful constraints for future population synthesis studies and provided motivation for future studies of similar LIRGs. Perhaps complex age effects implicated in this study can be teased out on short timescales using additional galaxies and subgalactic regions like in NGC~7552, such that HMXB emission levels could eventually serve as a reliable indicator of average population age.

\end{itemize}

\section{Acknowledgments}

We gratefully acknowledge support from FINESST award 19-ASTRO19-0120 and grant no.~80NSSC19K1415 (L.W.,B.D.L.), \chandra\ Grant No. GO8-19039X (B.D.L., A.P., R.E.), and \nustar\ Grant No. 80NSSC20K0030 (L.W., B.D.L.). K.G.'s research was supported by an appointment to the NASA Postdoctoral Program at NASA Goddard Space Flight Center, administered by Oak Ridge Associated Universities under contract with NASA. We thank Bernhard Brandl, Hsi-An Pan and Tony Wong for sharing data products and providing guidance on their use. We thank the referee for their helpful suggestions, which helped improve the quality of this paper.\\

\vspace{5mm}
\facilities{HST, CXO, NuSTAR}

\software{CIAO\citep[v4.8, v4.11;][]{Fruscione2006}, HEASoft \citep[v6.25;][]{NASA2014}, SAOImage DS9 \citep[v8.0;][]{Joye2003}, XSpec \citep[v12.10.0c;][]{Arnaud1996}.
          }
\software{This manuscript has made use of the following Python modules: 
\texttt{numpy} \citep{Harris2020},
\texttt{scipy} \citep{SciPy2020},
\texttt{pandas} \citep{McKinney2010},
\texttt{matplotlib} \citep{Hunter2007},
\texttt{astropy} \citep{2013A&A...558A..33A,2018AJ....156..123A}.
}

\bibliography{references}{}
\bibliographystyle{aasjournal}

\end{document}